\documentclass[aps,pre,showpacs,twocolumn,groupedaddress]{revtex4-1}

\usepackage{graphicx}
\usepackage{dcolumn}
\usepackage{bm}
\usepackage{enumerate}
\usepackage{amsmath}
\usepackage{footnote}

\begin{document}

\title{ \bf Mean Field and Mean Ensemble Frequencies of a System of Coupled Oscillators}

\author{Spase Petkoski$^1$}

\author{Dmytro Iatsenko$^1$}

\author{Lasko Basnarkov$^{2,3}$}

\author{Aneta Stefanovska$^1$}

\email[]{aneta@lancaster.ac.uk}
\affiliation{$^1$Department of Physics, Lancaster University, Lancaster LA1 4YB, United Kingdom}
\affiliation{$^2$Ss. Cyril and Methodius University, Faculty of Computer Science and Engineering, P.O. Box 393, Skopje, Macedonia}
\affiliation{$^3$Macedonian Academy of Sciences and Arts, Skopje, Macedonia}

\date{\today}

\begin{abstract}
We investigate interacting phase oscillators whose mean field is at a different frequency from the mean or mode of their natural frequencies. The associated asymmetries lead to a macroscopic travelling wave. We show that the mean ensemble frequency of such systems differs from their entrainment frequency. In some scenarios these frequencies take values that, counter-intuitively, lie beyond the limits of the natural frequencies. The results indicate that a clear distinction should be drawn between the two variables describing the macroscopic dynamics of cooperative systems. This has important implications for real systems where a non-trivial distribution of parameters is common.

\end{abstract}

\maketitle

\section{Introduction}
\label{sec:Intro}

Systems consisting of large numbers of interacting units are common in science and nature, and have been the essential modelling tools
in physics, biology, chemistry and social science \cite{Strogatz_sync}. Here, the state of the whole system is characterized with macroscopic variables such as the temperature, magnetization and so on. The values of these depend on both the microscopic laws governing the dynamics of the units as well as their interaction.
Generally, only macroscopic variables are accessible in experiments. Thus their precise definition and interpretation is needed.

In the case of populations of weakly interacting oscillators, application of the phase approximation leads to the Kuramoto model (KM) for globally coupled phase oscillators  \cite{Kuramoto_book}. Although the model itself represents an idealized scenario, its analytical tractability makes it the prevailing  approach in tackling a wide variety of important problems --  from Josephson-junction  arrays \cite{Wiesenfeld1996} to brain dynamics under anaesthesia  \cite{Jane1} and  pedestrian induced oscillations in the Millenium bridge problem \cite{Strogatz2005}. This has led to many extensions of the basic model to allow more realistic descriptions of actual systems, e.g. KM under influence of external fields  \cite{Kuramoto1988}, or with time-varying parameters \cite{Petkoski:12} (for a review of generalizations and the problems they address see \cite{Acebron_review} and references therein).

A fundamental feature of this model is that for a large enough coupling, synchronized behaviour emerges. Depending on their inherent frequencies, some of the oscillators become locked, while the others continue to rotate asynchronously but with adjusted frequencies. The degree of the synchronization is usually characterized by some order parameter.
For example, in the paradigmatic example of flashing fireflies  \cite{Strogatz_sync}, this parameter would describe the fraction of fireflies that flash in synchrony.

Since the building units of the KM are defined by their natural frequencies, the macroscopic dynamics of the oscillating system must be also characterized by some average frequency.
However, two  quantities can be used:
the effective frequency to which synchronized oscillators are locked and the average frequency of all the oscillators, locked and unlocked, that belong to the observed system.
The former represents the natural macroscopic frequency, whilst the latter is the microscopically averaged mean frequency.
We shall call these respectively the {\it mean field frequency} and the {\it mean ensemble frequency}.
Returning to the example of fireflies, the frequency of those flashing in synchrony --  the mean field frequency -- can generally deviate from the observed mean frequency of the whole population -- the mean ensemble frequency. Similarly, some of the neurons in the brain are expected to be mutually locked to a certain frequency, whereas an electroencephalographic recording contains the mean of all neurons in some area, not just the synchronized.


In general, there is no reason for these frequency definitions to coincide. Still, not enough attention has been paid in formulating them for different parameters. Namely, due to the equality of the frequencies in the cases with symmetry that were mostly studied, they were used interchangeably and without verification, even when they do differ (e.g. see \cite{Strogatz2011}). However, here we consider scenarios which most closely resemble the actual physical or natural phenomena;
in particular, models with asymmetrically distributed frequencies, phase shifted coupling function, or  asymmetric couplings of opposite sign.
For them we show that these frequencies always differ and  have non-trivial values.
Hence, one should be extremely cautious when the measured frequency of a population is interpreted and then compared  with the theoretical model.

We begin with a formulation of the model and its group dynamics parameters. Section \ref{sec:Stat} describes the stationary solutions of the KM, whilst all possible scenarios with non-trivial  mean field and mean ensemble frequencies are described  in Section \ref{sec:TW}.
The summary of the work and its implications are discussed in Section \ref{sec:disc}.

\section{Formulation}
\label{sec:Formulation}

The KM consists of phase oscillators running at arbitrary intrinsic frequencies and coupled through the sine of their phase differences.
In the case of  heterogenous coupling strengths, the dynamics of the phase $\tilde{\theta}_i$ of the $i$th oscillator has the form
\begin{eqnarray}
\dot{\tilde{\theta}}_i =\tilde{\omega}_i +\frac{K_i}{N}\sum_{j=1}^{N} \sin(\tilde{\theta}_j - \tilde{\theta}_i), \ \ \ i=1,\ldots,N.
 \label{eqn:KM}
\end{eqnarray}
Here,   $K_i$ is the coupling strength of each oscillator and it is drawn from a probability distribution $\Gamma(K)$.
Similarly, the natural frequencies $\tilde{\omega}_i$ are randomly distributed according to some  $\tilde{g}(\tilde{\omega})$.
The tildes  for the frequencies, phases and their distributions later are intentionally used for reasons to be explained below.
Without loss of generality, we assume $\tilde{g}(\tilde{\omega})$ to have a mean  $<\tilde{\omega}_i>$ centered at $0$.
Hereafter this frame of reference will be called natural. Kuramoto introduced a complex order parameter for this model, defined as a centroid of the complex representation of the oscillators
\begin{eqnarray}
\tilde{z}(t)\equiv r(t)\mathrm{e}^{i\tilde{\psi}(t)}=\frac{1}{N}\sum_{j=1}^{N} \mathrm{e}^{i\tilde{\theta}_j}.
 \label{eqn:z}
\end{eqnarray}
It characterizes the macroscopic behaviour of the oscillators, with $r$ and $\tilde{\psi}$ being the amplitude and the phase of the mean field respectively.
The former shows the level of synchronization, while the latter gives the position of the peak in the distribution of phases. Generally, the long-term values for both can depend on time. Applying Eq.~(\ref{eqn:z}) to the governing equation (\ref{eqn:KM}), it is rewritten as
\begin{eqnarray}
\dot{\tilde{\theta}}_i = \tilde{\omega}_i -\, K_i\, r\sin(\tilde{\theta}_i - \tilde{\psi}).
 \label{eqn:Kuramoto2}
\end{eqnarray}

For infinitely large populations $N \rightarrow \infty$, a probability distribution function (PDF)  $\tilde{f}(\tilde{\theta}, \tilde{\omega}, K, t)$ is defined, such that $\int_{-\pi}^{+\pi} \tilde{f}(\tilde{\theta}, \tilde{\omega}, K, t) \ d\tilde{\theta}  = g(\tilde{\omega}) \Gamma(K). $
Thus, the complex mean field Eq.~(\ref{eqn:z}) becomes
\begin{eqnarray}
\tilde{z} =  \int_{-\pi}^{\pi} \int_{-\infty}^{+\infty} \int_{-\infty}^{+\infty}  \mathrm{e}^{i \tilde{\theta}} \tilde{f}( \tilde{\theta}, \tilde{\omega}, K,t) \  d\tilde{\theta} \ d\tilde{\omega} \ d K.
\label{eqn:MFIntegral}
\end{eqnarray}
For convenience, infinite limits in all further definite integrals will be omitted.

As a consequence of the conservation of the number of oscillators the evolution of the density function is governed by a continuity equation
\begin{equation}
\label{eqn:continuity}
\frac{\partial \tilde{f}}{\partial t}=-\frac{\partial}{\partial \theta}\{[\tilde{\omega} +\frac{K}{2i}(\tilde{z}\mathrm{e}^{-i\tilde{\theta}}-\tilde{z}^\ast \mathrm{e}^{i\tilde{\theta}})]\tilde{f}\}. \end{equation}
Here the  right hand side of Eq.~(\ref{eqn:Kuramoto2}) is used and the sine function is  expressed  with complex exponents. Moreover, since $\tilde{f}$ is $2\pi$ periodic in $\tilde{\theta}$ it allows a Fourier expansion and can be written as
\begin{eqnarray}
\tilde{f} = \frac{\tilde{g}(\tilde{\omega}) \Gamma(K)}{2 \pi} \{1 + \sum_{k=1}^{\infty} [\tilde{f}_k (\tilde{\omega}, K, t) \ e^{i k \tilde{\theta}} + c.c.] \},
\label{eqn:fFourier}
\end{eqnarray}
where c.c. are the complex conjugates and $\tilde{f}_{-k}=\tilde{f}_k^{\ast}$.

In the limit $t \rightarrow \infty$, the ensemble described by (\ref{eqn:continuity}) might settle into a stationary state for some rotating frame.
We define: i) systems that have stationary solutions in some  frame of reference i.e. the complex mean field and the distribution of phases rotate uniformly and they have a constant mean field after the initial transitions; and ii) systems that experience complex non-stable behavior, i.e. a time-varying amplitude of the order parameter $r(t)$. These definitions might differ from usual descriptions found elsewhere which regard as stationary only those solutions that are fixed in the natural reference frame. In this work, our attention is focused on the ensembles with stationary solutions as described by i).

For the case of identically coupled oscillators with unimodal and symmetric distributions of their natural frequencies,
above the critical coupling a  phase locking of the oscillators takes place around the peak of $\tilde{g}(\tilde{\omega})$ where the density of the oscillators is highest \cite{Kuramoto_book}.
As a consequence of the symmetry, the group dynamics is stationary and both mean frequencies are equal to the mode of $\tilde{g}(\tilde{\omega})$, which in this case is also its mean value.

Nevertheless, introducing multimodal or asymmetric $\tilde{g}(\tilde{\omega})$, or distributed $K$ leads to much richer dynamics.
Thus,  for  multimodal $\tilde{g}(\tilde{\omega})$  bistabilities and standing waves  emerge \cite{Acebron1998D, Acebron1998pre, Montbrio2004}.
A standing wave is a macroscopic solution where neither  $\tilde{f}(\tilde{\theta}, \tilde{\omega}, K, t)$, nor   $\tilde{z}(t)$ are stationary in any rotating frame.
This further implies non-stationarity of $r(t)$.
Standing waves  are also observed in systems with symmetrical bimodal distribution of natural frequencies,  with the exact result for bifurcations between different states given in \cite{Strogatz2009}.

Travelling waves (TW), as another peculiar group behaviour, have also been observed for different parameter ranges within the same systems  \cite{Acebron1998D, Acebron1998pre}.
In our analysis, a TW state is considered to be any solution characterized by long-term stationarity of the mean field amplitude, whereas the frequency of the locking $\Omega$ differs from the mean of the natural frequencies.
In other words, the locking of synchronized oscillators is in a frame different from the natural.
This also represents a stationary solution according to its definition above.

A recent study \cite{Strogatz2011} shows the occurrence of TW in models with positive and negative coupling strengths, and identifies so called conformists  and contrarians. Similarly, a synchronization around a frequency that is different from the mean or the peak of the distribution was reported for ensembles that have an asymmetric unimodal distribution of natural frequencies \cite{Lasko2008}. It is also worth mentioning here the Kuramoto-Sakaguchi model \cite{Sakaguchi1986}, where the  phase shift  is  introduced into the coupling function, to allow synchronization at a frequency different from the mean of the natural frequencies.
Hence, it always leads to TW states. Additionally, the whole class of models of non-isochronous oscillators with constant shear can be reduced to this model \cite{Kuramoto_book, Montbrio2011}.

Stationary solutions for fully symmetric populations  have macroscopic frequencies that are equal to $<\tilde{\omega}_i>$.
In asymmetric scenarios on the other hand, the synchronized cluster experience non-trivial phase velocity \cite{Sakaguchi1986, Lasko2008, Strogatz2011}.
Thereafter, the focus of this work is interacting phase oscillators whose coherent behavior is characterized by a TW state, as defined earlier.
Having certain asymmetries, either in the frequencies, the coupling parameters, or in the coupling function itself, is a necessary condition for occurrence of this state.
As a consequence, the influence of the unsynchronized oscillators to the entrainment frequency  does not vanish \cite{Sakaguchi1986, Lasko2008, Strogatz2011}.
Additionally, we show that these oscillators also cause the mean ensemble frequency to have a non-trivial value.
It generally differs from  the entrainment frequency, but also from the mean and the mode of the natural frequencies.

Before proceeding with the analysis of all cases with TW state, let us define the macroscopic frequencies that we have already discussed.
The mean field frequency represents the  velocity of the mean phase $\tilde{\psi}$  and is obtained from the time derivative of the complex mean field (\ref{eqn:z})
\begin{eqnarray}
\dot{\tilde{z}} e^{-i \tilde{\psi}} = \dot{r} + i\dot{\tilde{\psi}}r = \frac{1}{N} \sum_{j=1}^N  i \dot{\tilde{\theta}}_j e^{i (\tilde{\theta}_j - \tilde{\psi})}.
\label{eqn:OrderParameterDot2}
\end{eqnarray}
Taking into account that $r$ and $\psi$ are both real, the same is true for their time-derivatives, so from Eqs.~(\ref{eqn:Kuramoto2}) and (\ref{eqn:OrderParameterDot2}) one finally obtains following evolutions of the amplitude and the phase of the complex mean field
\begin{eqnarray}
\dot{\tilde{\psi}} \equiv \Omega =  \frac{1}{r N} \sum_{j=1}^N [\tilde{\omega}_j -\, K_j\, r\sin(\tilde{\theta}_j - \tilde{\psi})]  \cos(\tilde{\theta}_j - \tilde{\psi}). \ \
\label{eqn:OrderParameterDot3}
\end{eqnarray}
The expression for $\dot{\tilde{\psi}}$ represents the velocity of the mean phase, i.e. the frequency of the synchronized oscillators.
For the other frequency parameter -- the mean frequency of the ensemble -- used for characterizing these systems, its definition leads to
\begin{eqnarray}
\tilde{f}_{ens} \equiv \frac{1}{N} \sum_{j=1}^N  \dot{\tilde{\theta}}_j = \frac{1}{N} \sum_{j=1}^N [\tilde{\omega}_j -\, K_j\, r\sin(\tilde{\theta}_j - \tilde{\psi})].
\label{eqn:MFr}
\end{eqnarray}
In the infinite limit,  (\ref{eqn:fFourier}) is introduced into (\ref{eqn:MFIntegral}).
By taking the time derivative and applying the substitution (\ref{eqn:continuity}), the evolution of the complex order parameter is obtained
\begin{eqnarray}
\dot{\tilde{z}}
 = \int \int [i\tilde{\omega} \tilde{f}_1^{\ast}+\frac{K}{2}(\tilde{z}-\tilde{z}^{\ast}\tilde{f}_2^{\ast})] \tilde{g}(\tilde{\omega})\Gamma(K) d\tilde{\omega} dK. \
\label{eqn:MFIntegral3}
\end{eqnarray}
Similarly,  the mean frequency of the ensemble becomes
\begin{equation}
\tilde{f}_{ens} = \int_{-\pi}^{\pi} \int \int  \dot{\tilde{\theta}} \ \tilde{f}(\tilde{\theta}, \tilde{\omega}, K, t)  d\tilde{\theta}  d\tilde{\omega} d K.
\label{eqn:MFr2}
\end{equation}

\section{Stationary solutions for the phase distribution}
\label{sec:Stat}

For a large class of problems, the long term coherent dynamics of an ensemble of phase oscillators is  time-independent, as a consequence of the stationary distribution of the phases.
This implies existence of the stationary solution of the continuity equation (\ref{eqn:continuity}), which does not need to be in the natural reference frame.
A recent work \cite{Iatsenko2013} discusses generalized empirical stability conditions for these systems.

From now on we consider that the ensemble has non-zero mean field, i.e. it is out of the incoherent state (a fully incoherent solution is also stationary), and that stationary solutions for the phase distribution exist in some planes of reference. This is true for the simplest possible scenario -- unimodal symmetric frequency distribution and constant coupling strengths. The TW state is another possible scenario with this property,  despite the fact that in this situation the velocity of the mean phase (or the mean field frequency as defined here) differs from the mean of the natural frequencies.

In our analysis  the mean field frequency is allowed to be nonzero despite assuming that $<\tilde{\omega}> =0$. Still, we work in the frame where $\tilde{f}(\tilde{\theta}, \tilde{\omega}, K, t)$ is stationary -- the reference frame rotating  with the frequency $\Omega$. Here the phases of the oscillators are $\theta=\tilde{\theta} - \Omega t$ and the phase corresponding to the complex order parameter is $\psi=\tilde{\psi} - \Omega t$. The distribution of the natural frequencies becomes $g(\omega)=\tilde{g}(\omega+\Omega)$ with mean $-\Omega$. In the same frame $\psi =0$ can be assumed  after an appropriate phase shift.

For any ensemble, if  the stationary solution of (\ref{eqn:continuity}) exists, then it exhibits two types of long-term behavior, depending on the size of $|\tilde{\omega}-\, \Omega|=|\omega|$ relative to $|K r|$ and to the sign of $K$. The oscillators with $|\omega|<|K r|$ approach a stable fixed point defined implicitly by
\begin{eqnarray}
\label{eqn:locked condition2}
  \theta = \left\{
  \begin{array}{l l}
     \arcsin \frac{\omega}{|K| r}, & \quad \text{if $K>0$ }\\
    \pi + \arcsin \frac{\omega}{|K| r}, & \quad \text{if $K<0$ .}\\
  \end{array} \right.
\end{eqnarray}
These oscillators are called \emph{locked} because they maintain constant phase difference, while rotating at frequency $\Omega$ in the original frame.
It is also assumed that synchronized oscillators with positive couplings have phases in the interval $(-\pi/2, \pi/2)$, while those with negative are in $(\pi/2,  3 \pi/2)$, since these are necessities for stable solutions of Eq.~(\ref{eqn:locked condition2}) \cite{Acebron_review}.
In contrast, the oscillators with $|\omega_i| >|K_i r|$ rotate in a non-uniform manner. As expected, the locked oscillators correspond to the center of $g(\omega)$ and the drifting oscillators correspond to the tails.

For the synchronized oscillators the stationary distribution of the phases becomes
\begin{eqnarray}
 &&  f_s(\theta,\omega,K,t) = \nonumber \\
  && \left\{
  \begin{array}{l l}
     g(\omega)\Gamma(K)\delta[\theta-\arcsin (\frac{\omega}{|K|r})], & \quad \text{if $K>0$ }\\
    g(\omega)\Gamma(K)\delta[\theta-\arcsin (\frac{\omega}{|K|r})-\pi], & \quad \text{if $K<0$}.\\
  \end{array}
   \right.
\label{eqn:Disribution_Locked}
\end{eqnarray}
Oscillators with frequencies in the interval $|\omega| > r |K| $ are out of synchrony with the mean phase.
Their stationary distribution is obtained from the continuity equation (\ref{eqn:continuity})
and the normalization condition of $f(\theta,\omega,K,t)$, such that
\begin{eqnarray}
 f_{as}(\theta,\omega,K, t)= g(\omega)\Gamma(K)\frac{\sqrt{\omega^2-(Kr)^2}}{2\pi|\omega-Kr\sin(\theta)|}.
\label{eqn:Disribution_Unlocked}
\end{eqnarray}
Hence, the real and imaginary parts of the complex mean field definition, Eq.~(\ref{eqn:MFIntegral}), become
\begin{eqnarray}
\label{eqn:MF_Real}
r = \int_{-\pi}^{\pi} \int  \int  \cos\theta f(\theta,\omega,K, t) \  d\theta \ d\omega \ d K, \\
0 = \int_{-\pi}^{\pi} \int  \int  \sin\theta f(\theta,\omega,K, t) \  d\theta \ d\omega \ d K.
\label{eqn:PhBalance}
\end{eqnarray}
The latter  is identified as the phase balance equation.

The distribution of the phases for synchronized oscillators, Eq.~(\ref{eqn:Disribution_Locked}),  for each $\omega$ implies $\pi$ difference between the phases of the synchronized clusters with couplings of opposite sign. Nevertheless, this holds for the distribution $f_s(\omega,K,\theta, t)$ only if $g(\omega)$ is symmetric.
This is not the case for the TW state, where even if the oscillators were symmetrically distributed in the natural frame, the symmetry would be broken when moving to the frame rotating with $\Omega$. Thereafter, the centroids of the phases will be shifted from the $\pi$ mutual distance, a phenomenon that was reported in \cite{Strogatz2011}.

In the rotating reference frame the locked oscillators are frozen, i.e. by definition $\dot{\theta}=\dot{\psi}=0$. Thus only the drifting ones need to be considered.
This can be applied in the expression for the mean frequency of the ensemble, Eq.~(\ref{eqn:MFr2}), which becomes (see  Appendix \ref{sec:app1})
\begin{eqnarray}
f_{ens} =  \int\int_{|Kr|}^{\infty} [g(\omega)-g(-\omega)] \Gamma(K) \sqrt{\omega^2-(Kr)^2}d\omega dK.  \nonumber\\
\label{eq:Ensamble_Mean_Final}
\end{eqnarray}
The definition (\ref{eq:Ensamble_Mean_Final}) is not restricted to any distributions of the natural frequencies and couplings.

Going back to the rotating frame where the original $\tilde{g}(\tilde{\omega})$ has zero mean, frequency parameters of the TW state are characterized by $\Omega$ and $\tilde{f}_{ens}$, where $$\tilde{f}_{ens}=f_{ens} + \Omega.$$
As a consequence  of (\ref{eqn:PhBalance}) $\Omega$ can equal 0 only if $g(\omega)$ is symmetric in the boundaries of the integral in (\ref{eq:Ensamble_Mean_Final}) \cite{Lasko2008}.
This means that for any zero centered distribution of the natural frequencies $\tilde{g}(\tilde{\omega})$ which is even in the intervals $\tilde{\omega}>|rK|$ (e.g. Lorentzian, Gaussian, etc) the function inside the integral is also even if $\Omega=0$.
Thus the integrals cancel each other and $\tilde{f}_{ens}=f_{ens}=\Omega=0$.
\\

\section{The travelling wave states}
\label{sec:TW}

In the following we discuss the possible scenarios that lead to the TW state for which we also obtain mean frequency parameters described earlier.

All further analyses are carried out in the rotating frame of the  entrainment frequency $\Omega$, such that $\dot{\psi}\equiv0$ and natural frequencies are distributed according to $\tilde{g}(\omega+\Omega)$, where $\tilde{g}(\tilde{\omega})$ has zero mean. However, for better clarity of the figures, frequencies in the examples with asymmetric $\tilde{g}(\tilde{\omega})$  are depicted as seen from the original distribution described in the captions.
In other words,  compared to the results from the analysis, the plots are shifted by the means of the given distributions $\tilde{g}(\tilde{\omega})$.

\subsection{Traveling waves in the Kuramoto model with contrarians and conformists}
\label{subsec:TWCon}

First we focus on the TW solution described in \cite{Strogatz2011}, which actually inspired this work.
The model shows resemblance to sociophysical models of opinion formation \cite{Lama} and is also a continuation of the KM with distributed positive couplings \cite{Paissan2007}.

The distribution of the couplings is
$$\Gamma(K)=(1-p)\delta(K-K_1)+p\delta(K-K_2),$$
where $K_1<0$ and $K_2>0$, and $p$ denotes the probability that a randomly chosen oscillator is a conformist, while $q=1-p$ is the probability that a random oscillator is a contrarian. The natural frequencies follow a zero centered Lorentzian distribution with half-width $\gamma$
$$\tilde{g}(\tilde{\omega})=\frac{\gamma}{\pi(\tilde{\omega}^2+\gamma^2)}.$$
As stated in \cite{Strogatz2011}, if the absolute coupling strength  is higher for  conformists than for the contrarians, then for some region in the parameters space $\gamma-p$, the synchronized oscillators will experience a TW. This means that both peaks in the phase distribution uniformly  rotate in same direction.
The waves appear in symmetric pairs, with frequencies $\pm \Omega$, since they result from the asymmetry in the coupling strengths.
On the contrary, if the TW is due to the asymmetry in the natural frequencies or in the coupling function, the waves are not paired, as discussed later.

Following the definitions  of $\Gamma(K)$ and $\tilde{g}(\tilde{\omega})$, and using the substitution  $\omega^2-(K_{1/2}r)^2=u_{1,2}^2$, the integrals in Eq.(\ref{eq:Ensamble_Mean_Final}) are analytically  solved, yielding
\begin{widetext}
\begin{eqnarray}
f_{ens}  &=& -\sqrt{2}\Omega \gamma  \Bigl\{ \frac{1-p}{\sqrt{\sqrt{(\gamma^2+\Omega^2 + K_1^2 r^2)^2 -4 K_1^2 r^2 \Omega^2}-\Omega^2+K_1^2 r^2+\gamma^2}}+  \nonumber \\
&+&  \frac{p}{\sqrt{\sqrt{(\gamma^2+\Omega^2 + K_2^2 r^2)^2 -4 K_2^2 r^2 \Omega^2}-\Omega^2+K_2^2r^2+\gamma^2}} \Bigr\}.
\label{eq:Ensamble_Mean_FinalTW5}
\end{eqnarray}
\end{widetext}
This  expression can be straightforwardly generalized for multimodal-$\delta$ distributed coupling strengths.

\begin{figure}[t!]
\centering
\includegraphics[width=0.47\textwidth]{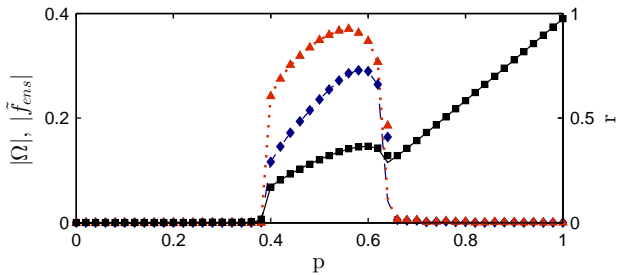}
\caption{(color online) The amplitude $r$ and frequency $\Omega$ of the mean field, and the mean ensemble frequency  $\tilde{f}_{ens}$ versus the ratio $p$.
Theoretical results are given with a solid line (black) for $r$, a dotted line (red)  for $\Omega$, and a dashed line (blue) for   $\tilde{f}_{ens}$.
Results from  numerical simulations are shown with squares (black) for $r$,  triangles (red) for $\Omega$, and diamonds (blue) for $\tilde{f}_{ens}$.
Parameters: $\gamma=0.05$,  $K_1=-1$ and $K_2=2$.
}
\label{fig:1}
\end{figure}

It is obvious that $\Omega=0$ will imply $\tilde{f}_{ens}=0$.
Hence only in the presence of a TW may the mean frequency of the ensemble differ from the mean phase velocity  for this model. Similarly, for the TW state
$f_{ens}$  is  non-zero and has opposite sign from $\Omega$. Additionally it can be shown that the expression in the curly brackets is  smaller than $1/\gamma$, so that
in the natural reference frame, $|\tilde{f}_{ens}|<|\Omega|$ always holds. This is also evident from the numerical results plotted in Fig.~\ref{fig:1}.

Let us now derive the expression for obtaining the frequency $\Omega$ of the TW, as seen in the natural reference frame.
Ott and Antonsen, in their seminal work \cite{Ott1}, showed that  macroscopic evolution of large systems of coupled oscillators can be described by an explicit definite set of nonlinear differential equations. They introduced an ansatz for the complex Fourier coefficients in Eq.~(\ref{eqn:fFourier})
\begin{equation}
\label{eqn:ansatz}
\tilde{f}_n(\tilde{\omega},K,t) = [\tilde{\alpha}(\tilde{\omega},K,t)]^n,
\end{equation}
which exactly solves the governing equation (\ref{eqn:continuity}), as long as $\tilde{\alpha}(\tilde{\omega}, K, t)$ evolves  following the nonlinear equation
\begin{eqnarray}
\label{eqn:ansatz2}
\frac{\partial \tilde{\alpha}}{\partial t} + i \tilde{\omega} \tilde{\alpha} + \frac{K}{2}(\tilde{z} \tilde{\alpha}^2 - \tilde{z}^{\ast}) = 0.
\end{eqnarray}
When this ansatz is implemented in Eq.(\ref{eqn:MFIntegral}), the order parameter reduces to
\begin{eqnarray}
\label{eqn:z*}
\tilde{z}=\int \int \tilde{\alpha}^\ast(\omega, K, t)  \tilde{g}(\tilde{\omega})\Gamma(K) d\tilde{\omega} dK.
\end{eqnarray}
Note that applying this ansatz in (\ref{eqn:MFIntegral3}), leads to a simplified expression for the evolution of the complex order parameter
\begin{eqnarray}
\dot{\tilde{z}}  =
\int \int [i\tilde{\omega} \tilde{\alpha}^{\ast}+\frac{K}{2}(\tilde{z}-\tilde{z}^{\ast}\tilde{\alpha}^{\ast 2})] \tilde{g}(\tilde{\omega})\Gamma(K) d\tilde{\omega} dK, \ \
\label{eqn:MFIntegral4}
\end{eqnarray}
from where  the frequency of entrainment can be obtained. Similarly, substituting Eqs.~(\ref{eqn:Kuramoto2}) and (\ref{eqn:fFourier}) and the ansatz (\ref{eqn:ansatz}) into Eq.~(\ref{eqn:MFr2}) transform the mean ensemble frequency to
\begin{eqnarray}
\tilde{f}_{ens}=\int \int [\tilde{\omega}-\frac{K}{2i}(\tilde{z}\alpha-\tilde{z}^{\ast}\tilde{\alpha}^{\ast})]\tilde{g}(\tilde{\omega})\Gamma(K)d\tilde{\omega} dK. \ \ \ \ \ \ \
\label{eqn:MFr4}
\end{eqnarray}

We start by considering the low-dimensional evolution (\ref{eqn:ansatz2}) in the reference frame of the TW. Similar analysis, but in the natural frame was performed in \cite{Strogatz2011}.
First  the bimodal-$\delta$ and Lorentzian distributions for $\Gamma(K)$ and $g(\omega)$ respectively, are substituted into the integral  Eq.~(\ref{eqn:z*}), such that it yields
\begin{eqnarray}
\label{eqn:TWOA2}
 z^\ast = (1-p) \alpha_{1}(\Omega-i\gamma, K_{1}) + p \alpha_{2}(\Omega-i\gamma, K_{2}).
\end{eqnarray}
This is then used in  Eq.~(\ref{eqn:ansatz2}), which rewritten for both, $\alpha_{1}(\Omega-i\gamma, K_{1})$ and $\alpha_{2}(\Omega-i\gamma, K_{2})$, results in
\begin{eqnarray}
\label{eqn:TWOA1}
\frac{\partial \alpha_{1,2}}{\partial t} &&+ (i\Omega + \gamma) \alpha_{1,2} + \frac{K_{1,2}}{2}(z \alpha_{1,2}^2 - z^\ast) = 0,
\end{eqnarray}
where we omit the dependencies  of $ \alpha(\omega, K, t)$.
As previously  stated, we are interested in a stationary solution of the TW state in the $t\rightarrow\infty$ limit.
This implies time-independent distribution of the phases in this limit, or  from the ansatz (\ref{eqn:ansatz}), time-independent $\alpha_{1,2}$ with $\partial \alpha_{1,2} / \partial t=0.$
Further,  similar to \cite{Strogatz2011}, complex order parameters for each of the subpopulations, and the difference between their phases are defined
\begin{eqnarray}
\label{eqn:def}
r_1 e^{-i \psi_1}=\alpha_1, \, \ r_2 e^{-i \psi_2}=\alpha_2, \, \ \delta=\psi_1-\psi_2 = {\rm const}.  \ \ \ \
\end{eqnarray}
In this way it is ensured that  both synchronized populations, in-phase and antiphase, rotate with the same velocity $\Omega$ in the natural frame, and preserve constant phase difference.
In the rotating frame of the TW $\dot{\psi}=0$, so $\psi\equiv0$ can be set without loss of generality.
Thus, from Eqs.~(\ref{eqn:TWOA1} - \ref{eqn:def}) we obtain the following evolutions which describe a fixed point in the $\{r_{1}, r_{2}, \psi_1, \psi_2\}$ space.
\begin{eqnarray}
\label{eqn:TWOAr1}
 && \dot{r}_1 =  - \gamma r_1 - \frac{K_1}{2}[(r_1^2-1) (p r_2 \cos \delta + q r_1)] = 0, \ \\
 \label{eqn:TWOAr2}
 && \dot{r}_2 =  - \gamma r_2 - \frac{K_2}{2}[(r_2^2-1) (p r_2 + q r_1 \cos \delta)] = 0, \ \\
  \label{eqn:TWOApsi1}
 && \dot{\psi}_1 =   \Omega - \frac{K_1}{2 r_1} p r_2 \sin \delta (r_1^2 + 1)  = 0, \ \\
  \label{eqn:TWOApsi2}
 && \dot{\psi}_2 =   \Omega + \frac{K_2}{2 r_2} q r_1 \sin \delta (r_2^2 + 1)  = 0.
\end{eqnarray}
The low-dimensional parameters including $\Omega$ can be now obtained self-consistently. The steady states (\ref{eqn:TWOApsi1}-\ref{eqn:TWOApsi2}) result from the constant angle difference between the peaks in the phase distribution, i.e.
\begin{eqnarray*}
   \label{eqn:TWOAdelta}
\dot{\delta}=-\sin\delta[\frac{K_1}{2r_1}pr_2(r_1^2+1)+\frac{K_2}{2r_2}q r_1(r_2^2+1)]=0.  \ \
\end{eqnarray*}
Hence, when $Im [z]=0$ is applied to  Eq.~(\ref{eqn:TWOA2}),  one can use the expression
\begin{eqnarray}
 q r_1 \sin \psi_1 = - p r_1 \sin \psi_2
 \label{eqn:TWOA6}
\end{eqnarray}
and the definition (\ref{eqn:def}) of $\delta$ to obtain the values of $\psi_1$ and $\psi_2$, such that the system will be fully described.

The equations (\ref{eqn:TWOAr1}-\ref{eqn:TWOApsi2}) and all further numerical integrations are performed using a Runge-Kutta 4th order algorithm.
Once we have the low-dimensional parameters, Eq.~(\ref{eq:Ensamble_Mean_FinalTW5}) is applied to find  the mean frequency of the ensemble.
Finally, results are compared with the values for the order parameter $r$, the mean phase velocity of the ensemble $\Omega$ and the mean frequency $f_{ens}$, obtained from the numerical simulations of the ensemble Eq.~(\ref{eqn:KM}), using  Eq.~(\ref{eqn:z}) and Eq.~(\ref{eqn:MFr}). As in all later simulated scenarios, the number of oscillators was set to $N=100000$, the time step of the integration was $0.01$.
The simulations were running for $10^5$ time steps, with the initial $90\%$ of each run discarded as possibly transient, while the rest were time averaged.
The proportion of the conformists $p$ is changed from $0$ to $1$ at $100$ equally spaced points, and the obtained results  given in Fig.~\ref{fig:1} fully confirm the theoretical analysis.

\subsection{Asymmetric unimodal frequency distribution}
\label{subsec:AsF}

Another case of the KM characterized with a stationary solution, where the mean frequency of locked oscillators differs from the mean of the oscillators' natural frequencies, are ensembles with asymmetric unimodal  distribution  of the   natural  frequencies, and equal couplings. The asymmetric scenario is also  more  natural than the symmetric, because  any  imperfection in the system,  however  small,  can  destroy  the ideal symmetry. Nevertheless, mostly due to the analytical difficulties, this case  has obtained little attention. It was first examined by Sakaguchi and Kuramoto \cite{Sakaguchi1986} and the self-consistent condition for the mean field frequency was obtained in \cite{Lasko2008}.

It is known from \cite{Lasko2008}, that the  phase balance  equation (\ref{eqn:PhBalance}) for  $g(\omega)$ asymmetric in the interval $|\omega|> r K$, implies that the oscillators  always lock to a frequency that differs from the mean of the natural frequencies.
For the mean ensemble frequency Eq.~(\ref{eq:Ensamble_Mean_Final}) was numerically integrated  for  ensembles with triangular and lognormal frequency distributions and constant coupling strength.
We found that it always gives $f_{ens}=-\Omega$ in the TW rotating frame.
In other words, the mean ensemble frequency equals the mean of the natural frequencies, although it differs from the mean field frequency.
The results  seemed unexpected at first sight.
However after careful analysis we realized that the derivations for the equation of balance Eq.~(\ref{eqn:PhBalance}), as given in \cite{Lasko2008},  in the frame rotating with  $\Omega$ indeed leads to the  Eq.~(\ref{eq:Ensamble_Mean_Final}).
Namely, applying the derivations in the Appendix \ref{sec:app1}, Eq.~(\ref{eqn:PhBalance}) becomes
\begin{eqnarray}
&& 0 = \Omega +  \int\int_{|Kr|}^{\infty}  g(\omega) \sqrt{\omega^2-(Kr)^2}d\omega \nonumber\\
&-& \int\int_{-\infty}^{-|Kr|} g(\omega) \sqrt{\omega^2-(Kr)^2} d\omega = \Omega + f_{ens}. \ \ \
\label{eq:Ensamble_Mean_Final_AS}
\end{eqnarray}
Thus, one can conclude  that for the KM with asymmetric frequency distribution, the mean ensemble frequency is always equal to the mean of the natural frequencies.
\begin{figure}[t!]
\centering
\includegraphics[width=0.47\textwidth]{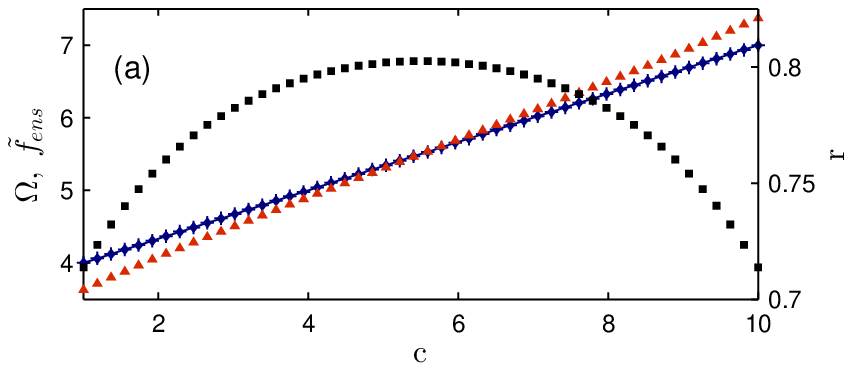}
\centering
\hspace*{-0.35cm}\includegraphics[width=0.47\textwidth]{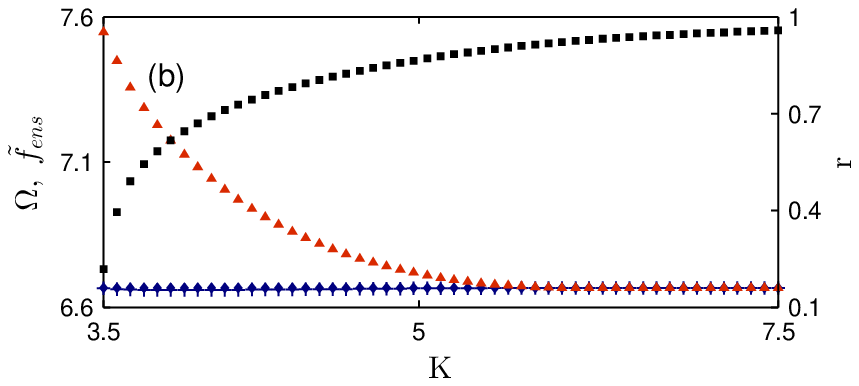}
\centering
\hspace*{-0.15cm}\includegraphics[width=0.46\textwidth]{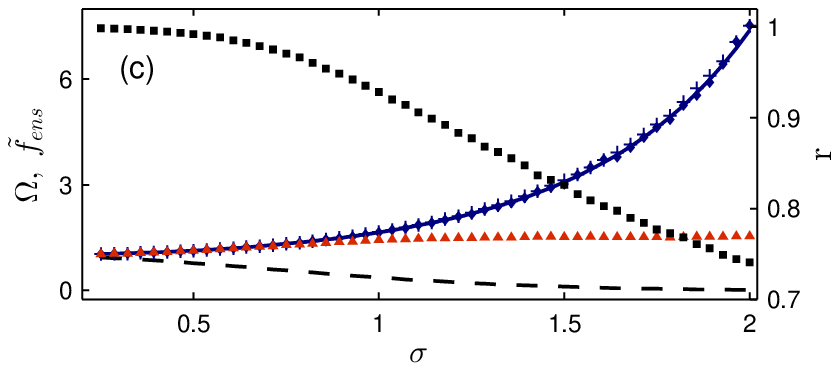}
\caption{(color online) The amplitude $r$ and frequency $\Omega$ of the mean field, and the mean ensemble frequency $\tilde{f}_{ens}$ for unimodal asymmetric natural frequencies.
Results from the numerical simulations are shown with  squares (black) for $r$,  triangles (red) for $\Omega$ and  diamonds (blue) for $\tilde{f}_{ens}$.
The  lines (blue) are from theoretical results for $\tilde{f}_{ens}$ and they also match the means of the frequencies' distributions -- crosses  (blue).
(a-b) Triangular  $\tilde{g}(\tilde{\omega})$ in boundaries $a = 1$ and $b = 10$.
(a) Coupling $K = 4.2$ and mode $c \in [0, 10]$;  (b) $K \in [4, 10]$ and $c = 9$;
(c) log-normal distribution of natural frequencies, with $\mu = 0$, $ \sigma \in [0.25, 2]$ and coupling $K = 4.2$. The dashed line shows the modes of the distributions.
}
\label{fig:2}
\end{figure}

Next, a triangular distribution with limits $a$ and $b$, and  peak at $c$ was explored.
In the scenario shown in Fig.~\ref{fig:2} (a) the peak is distributed
in the interval $[a, b]$, while in Fig.~\ref{fig:2} (b) the coupling strength is increasing  from $4$ to $10$  for fixed $c$.
Similarly, for log-normal frequencies
$$\tilde{g}(\tilde{\omega})=\frac{1}{\tilde{\omega} \sigma \sqrt{2 \pi}} e^{-(\ln \tilde{\omega} -\mu)^2/(2 \sigma^2)}, \ \ \tilde{\omega}>0,$$
$\mu$ is fixed to $0$, while $\sigma$ is logarithmically distributed, and the results are given in Fig.~\ref{fig:2} (b).
The mean ensemble frequency values also match theoretical values for the means of the given distributions, $(a+b+c)/3$ and $e ^{\mu +\sigma^2/2}$ respectively.

For the triangular distribution shown in Fig.~\ref{fig:2} (a) the mode is $c$, while for  (b) is fixed. So for the log-normal $\tilde{g}(\tilde{\omega})$ given in Fig.~\ref{fig:2} (b) the mode is $1$ for $\mu=0$ and $\sigma\rightarrow0$ and then exponentially decreases to $0$ with $e^{\mu-\sigma^2}$ as shown. Hence, the presented results show that the mean field frequency is always between the mode and the mean of the  distribution of natural frequencies, and reaches the second only when all  oscillators become synchronized. That is to say, by increasing the coupling strength for a given frequency distribution, the proportion of synchronized oscillators is also increased. Accordingly the value of $\Omega$ moves closer to the mean of the distribution, until it eventually reaches it for $r\rightarrow1$. For unbounded $g(\omega)$ the last can only occur when $K\rightarrow\infty$, while in the case of bounded natural frequencies, for some value of $r K$ all oscillators will be entrained. This can also be deduced from  Eq.~(\ref{eq:Ensamble_Mean_Final_AS}). Namely, higher synchronization implies a smaller region for the integral on the right hand side. At the same time we assume $g(\omega)$ to be decreasing left and right from the mode, meaning that higher $r$ leads to smaller value of $f_{ens}$ which eventually becomes $0$, implying $\Omega=\tilde{f}_{ens}$. Furthermore, the value under the square root in the same integral also decreases with increasing $r$. These are confirmed in Fig.~\ref{fig:2} (b-c), where we see that for smaller couplings and hence for smaller mean field amplitudes, $\Omega$ is closer to the peak of the distribution and approaches $\tilde{f}_{ens}$ for larger coupling.

\subsection{Asymmetric multimodal frequency distribution}
\label{subsec:AsBF}

The KM with multimodal asymmetric distribution of natural frequencies is another candidate for the mean field behavior described by a TW.
Nevertheless, due  to  difficulties that arise in the mathematical analysis of this model, it  was  never  fully solved, nor has a thorough  dynamical analysis of  possible macroscopic solutions  been performed.
Still, following the analysis of the symmetric scenario  \cite{Strogatz2009} and qualitative descriptions of the dynamics in the asymmetric case given in \cite{Kuramoto_book}  and \cite{Strogatz2009}, some conclusions can be drawn.
Namely, Kuramoto, in his seminal work  \cite{Kuramoto_book}, discusses  how transition from incoherence to mutual synchronization might be modified when the oscillators' natural frequencies  are  bimodally  distributed.
For sufficiently large coupling strength, he assumed that the clusters of synchronized oscillators
``will eventually be entrained to each other to form a single giant oscillator''.
For smaller coupling strengths compared to the distance between peaks, he envisaged that the synchronized nuclei would be at the peaks of $g(\omega)$.
Although some of the transitions  between different states described in  \cite{Kuramoto_book} for the symmetric case were shown to be wrong \cite{Strogatz2009}, the description of the mean field dynamics of a partially synchronized state is indeed correct.

Hence, for a symmetrical $g(\omega)$ the ensemble should be either partially synchronized for large enough coupling compared to the peaks' distance, or  synchronized clusters should exist near the peaks \cite{Kuramoto_book,Strogatz2009,Pazo2009}.
For asymmetrically and bimodally distributed frequencies the partial synchronization will be characterized by TW,  whereas standing waves and other complex
collective rhythms appear for smaller couplings.

\begin{figure}[t!]
\centering
\includegraphics[width=0.47\textwidth]{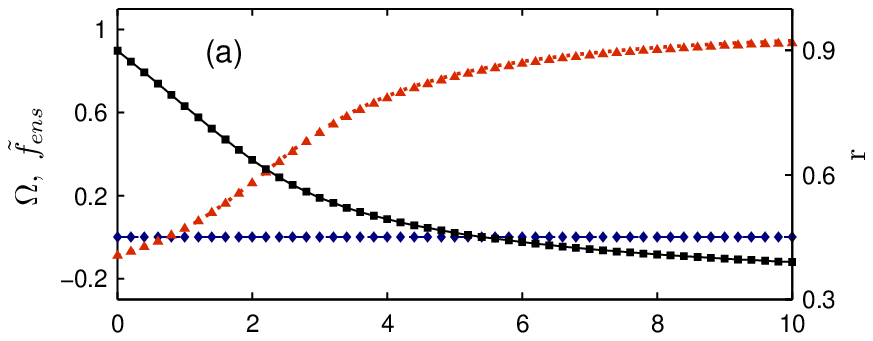}
\centering
\includegraphics[width=0.47\textwidth]{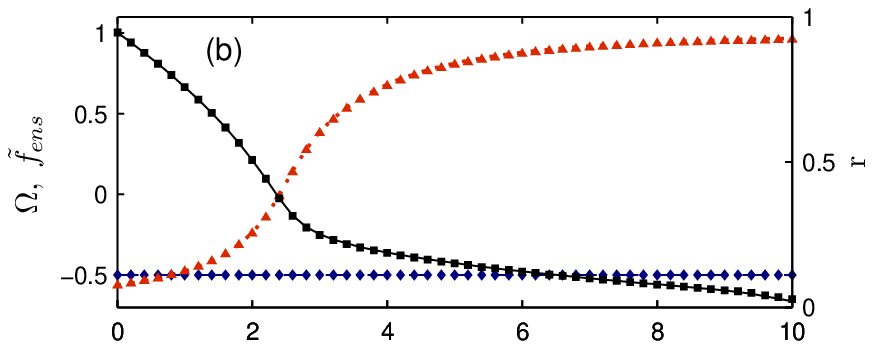}
\centering
\hspace*{0.15cm} \includegraphics[width=0.47\textwidth]{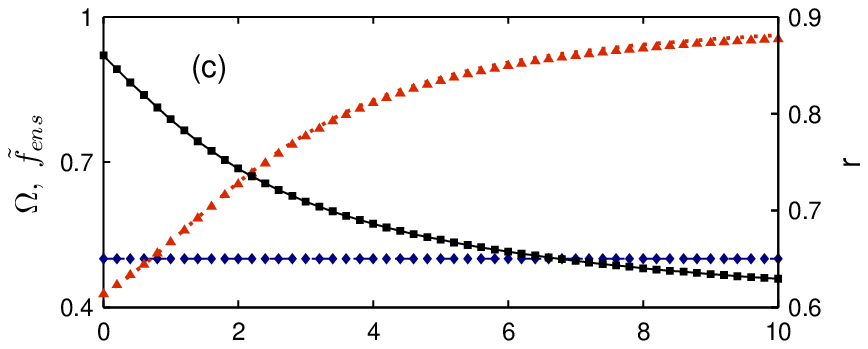}
\centering
\includegraphics[width=0.47\textwidth]{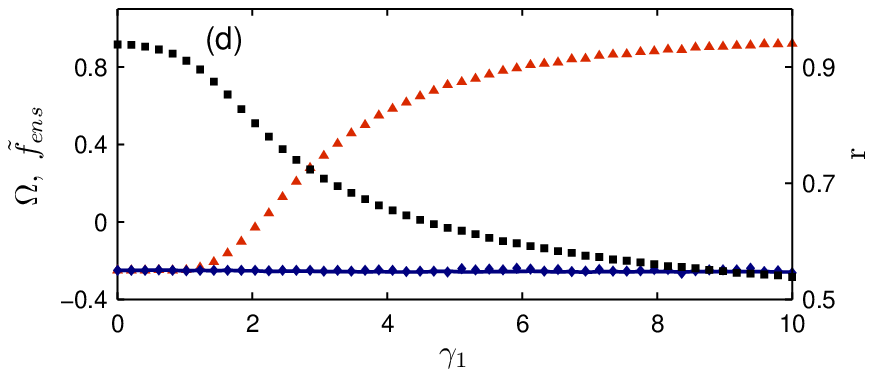}
\caption{(color online)  The amplitude $r$ and frequency $\Omega$ of the mean field, and the mean ensemble frequency  $\tilde{f}_{ens}$ versus the frequency width of the first subpopulation $\gamma_1$.
(a-d) Results from the simulations, Eq.~(\ref{eqn:KM}), are shown with  squares (black) for $r$, triangles (red) for $\Omega$ and diamonds (blue) for $\tilde{f}_{ens}$.
(a-c) The low-dimensional dynamics, Eq.~(\ref{eqn:BAS1}), are shown with solid lines  (black) for $r$, dotted lines (red) for $\Omega$ and dashed lines (blue)  for $\tilde{f}_{ens}$.
(a-c) Bimodal Lorentzian $\tilde{g}(\tilde{\omega})$. Parameters: $\gamma_1\in [0, 10]$ $\gamma_2=0.75$, $\mu_1=-1$, $\mu_2=1$, $K=5$, and (a) $p=0.5$, (b) $p=0.25$ and (c) $p=0.75$.
(d) Bimodal Gaussian $\tilde{g}(\tilde{\omega})$ with $\gamma_1\in [0, 10]$ $\gamma_2=0.75$, $\mu_1=-1.5$, $\mu_2=1$, coupling $K=4.25$ and $p=0.5$.
}
\label{fig:3}
\end{figure}

Thereafter, for the partially synchronized state in the asymmetric bimodal case, one might expect a TW state to occur.
As a consequence of having a single synchronized cluster, the dynamic of the system is similar to the models with unimodal asymmetric distribution described in Section \ref{subsec:AsF}.
As a result the same conclusion drawn about the asymmetric unimodal scenario, also holds for the TW solution of the asymmetric bimodal case.
This means that the mean frequency of the ensemble is equal to the mean of $\tilde{g}(\tilde{\omega})$, and as such differs from the frequency of the TW  experienced by the mean field.
Hence, in the TW frame, $f_{ens} = - \Omega$.
This can be seen in Fig. \ref{fig:3}, where results from numerical simulations confirmed the theoretically expected values for  $\tilde{f}_{ens}$ and $\Omega$.

As an example we analyze the dynamics of a population with bimodal Lorentzian distribution of frequencies
$$\tilde{g}(\tilde{\omega})= \frac{(1-p) \gamma_1}{\pi[(\tilde{\omega}-\mu_1)^2+\gamma_1^2]} +\frac{ p \gamma_2}{\pi[(\tilde{\omega}-\mu_2)^2+\gamma_2^2]}.$$
The subdistributions are peaked at $\mu_1$ and $\mu_2$ with the half-widths $\gamma_1$ and $\gamma_2$ respectively, where $p$ is the proportion of the oscillators belonging to the second subdistribution.
The full low-dimensional dynamics for this system  can be obtained using the ansatz Eq.~(\ref{eqn:ansatz}) in a similar manner as  in Section \ref{subsec:TWCon}.
The integral (\ref{eqn:z*}) for the given $\Gamma(K)$ and $g(\omega)$  yields
$$z^\ast = (1-p) \alpha_1(\mu_1-\gamma_1, K, t) + p \alpha_2(\mu_1-\gamma_1, K, t).$$
Hence, substituting $\alpha_{1,2}$ in Eq.~(\ref{eqn:ansatz2}), the low-dimensional dynamics are described by two ODEs
\begin{eqnarray}
\label{eqn:BAS1}
 \dot{\alpha}_{1,2} =  -(i \ \mu_{1,2} + \gamma_{1,2}) \alpha_{1,2} - \frac{K}{2}(z^\ast + z \alpha_{1,2}). \
\end{eqnarray}

In all the plots in Fig. \ref{fig:3}  the values of the coupling strength are large enough to induce TW  instead of standing wave states, i.e. one instead of two synchronized clusters. For the Fig. \ref{fig:3} (a) both peaks are of equal distance from the zero. This leads to  zero mean for any width of the subdistributions and following previous discussion and Eq.~(\ref{eq:Ensamble_Mean_Final_AS}) it implies $f_{ens}=0$. For similar deviations of subdistributions the entrainment frequency will be near zero, but closer to the narrower peak, since more oscillators from that subdistribution will be entrained. However, for a  large deviation of the first subdistribution compared to the second one, a  small number of the oscillators belonging to it will be entrained. These oscillators will be almost equally distributed on both sides of the second peak. Thus, the frequency of the entrained oscillators will asymptotically reach the peak of the second subdistribution
in the limit case $\gamma_1 \rightarrow \infty$. If in the same limit case $p \rightarrow 0$ also holds, then with the same reasoning very few entrained oscillators will be close to the second peak. At the same time the mean frequency of the distribution and of the ensemble will be near the first peak.
Similarly, for $p=0.25$ and $p=0.75$ as in Fig. \ref{fig:3} (b-c), the mean of the frequencies is on $1/4$ or $3/4$ of the distance between the peaks, respectively, while $\Omega$ reaches the narrower mode.

The generality of the described analysis for any bimodal distribution is shown in Fig. \ref{fig:3} (d), where the frequency distributions are Gaussian.
Although the low-dimensional dynamics in  a form similar to Eq.~(\ref{eqn:BAS1}) cannot be obtained for this case, the results from the numerical simulations of the ensemble, Eq.~(\ref{eqn:KM}), are in line with the previous discussion concerning Lorentizan $\tilde{g}(\tilde{\omega})$.

\subsection{Contrarians and conformists with asymmetric unimodal frequency distribution}
\label{subsec:AsFTW}

The analysis naturally continues with the KM in the presence of both previously analyzed conditions that are required for TW -- contrarians and conformists, with asymmetric distribution of natural frequencies. The phase locking in this scenario happens in frames other than the natural for all coherent solutions.
But as it will be shown, due to the distributed couplings, the mean ensemble frequency also has non-trivial values -- it is not always equal to $<\tilde{\omega}>$.
As another consequence of the asymmetry, the sign of the TW can no longer be expected to be random, as is the case for the contrarians and conformists over symmetrically distributed frequencies \cite{Strogatz2011}. However, the analysis of this case is largely  more complicated compared to previous ones and we have taken some examples only to show the main points.

\begin{figure}[t!]
\centering
\includegraphics[width=0.47\textwidth]{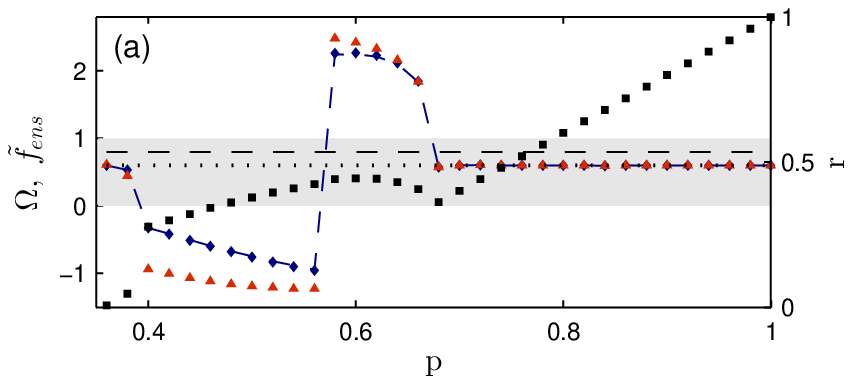}
\centering
\hspace*{0.15cm}\includegraphics[width=0.465\textwidth]{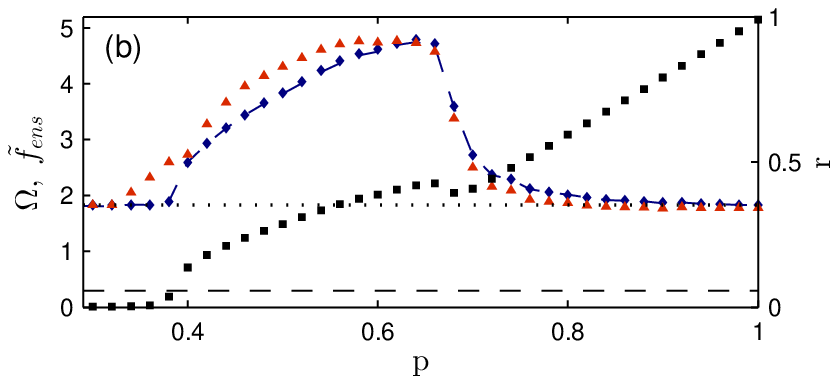}
\centering
\includegraphics[width=0.48\textwidth]{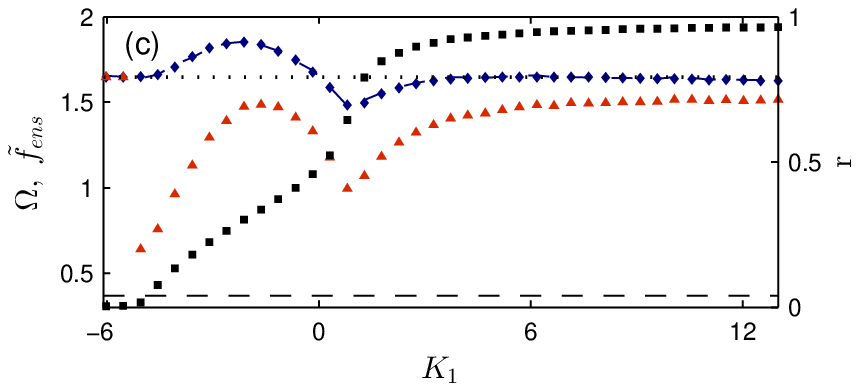}
\caption{(color online)  The amplitude $r$ and frequency $\Omega$ of the mean field, and the mean ensemble frequency  $\tilde{f}_{ens}$ for asymmetric ensembles with distributed coupling strengths.
Numerically obtained results, Eq.~(\ref{eqn:KM}), are shown with  squares (black) for $r$,  triangles (red) for $\Omega$ and diamonds (blue) for $\tilde{f}_{ens}$.
(a-c) The dashed  lines (blue) are theoretically predicted results for  $\tilde{f}_{ens}$, Eq.~(\ref{eq:Ensamble_Mean_Final}).
Horizontal dashed and dotted lines are the modes and the means of $\tilde{g}(\tilde{\omega})$ respectively, and its domain is shaded for (a).
(a) Triangular $\tilde{g}(\tilde{\omega})$  within $a = 0$ and $b = 1$ and peak at $c = 0.8$;  bimodal-$\delta$ $\Gamma(K)$ with $K_1=-4$, $K_2=8$ and $p\in [0.34, 1]$.
(b-c) Log-normal $\tilde{g}(\tilde{\omega})$  with $\mu = 0$ and bimodal-$\delta$  $\Gamma(K)$. (b) $ \sigma =1.1$, $K_1 = -10$, $K_2 = 20$; and (c) $ \sigma =1$, $p=0.55$, $K_1 \in [-6, 13]$, $K_2 = 6$.
}
\label{fig:4}
\end{figure}

The phase balance equation (\ref{eqn:PhBalance}) for distributed $K$ in this case does not lead to a simple expression as was Eq.~(\ref{eq:Ensamble_Mean_Final_AS}) for equal couplings. Still, following the same procedure, Eqs.~(\ref{eqn:Disribution_Locked}, \ref{eqn:Disribution_Unlocked}) are substituted into  Eq.~(\ref{eqn:PhBalance}) and applying the derivations from  Appendix \ref{sec:app1}, the balance of phases becomes
\begin{eqnarray}
&& 0 =  \int  \frac{\Gamma(K)}{r K} d K \{  \int \omega g(\omega)  d\omega \nonumber\\
&-& \int_{|Kr|}^{\infty}  \omega [g(\omega)-g(-\omega)] \sqrt{\omega^2-(Kr)^2} d\omega \} \nonumber\\
&=&  \int  \frac{\Gamma(K)}{r K} d K \ [-\Omega + I(K)].
\label{eq:Ensamble_Mean_Final_mix}
\end{eqnarray}
This expression self-consistently gives the frequency of the entrainment, whilst it no longer implies equality of $\tilde{f}_{ens}$ and $<\tilde{\omega}>$, as  for constant coupling parameters. Results from numerical simulations shown in Fig. \ref{fig:4} confirm the  non-trivial nature of both global frequency parameters.
The most interesting phenomena is that  $\Omega$ and $\tilde{f}_{ens}$ for some  parameters can have values that not only differ from either the mode and the mean of the natural frequencies, but that are also outside of the interval between them as was case for equal $K$. Moreover, they can even be outside of the boundaries of their distribution. For example for the triangular $\tilde{g}(\tilde{\omega})\in [0, 1]$ depicted in Fig.~\ref{fig:4} (a), the frequency of the entrainment can be up to $\sim -1$ or to $\sim 2$, and similar happens  with $\tilde{f}_{ens}$. These values are clearly out of the region of support of $\tilde{\omega}$, which is shaded in the same plot.
Note that these on first sight counterintuitive results are a consequence of the very high coupling strengths with opposite sign, and they are still within the general boundaries of $ \tilde{f}_{ens}$ and $\Omega$. Namely, from  Eqs.~(\ref{eqn:MFr2}) and (\ref{eqn:Kuramoto2}), it is clear that
\begin{eqnarray}
\hat{\omega} - \max|K| < \tilde{f}_{ens} < \hat{\omega} + \max|K|,
\label{eq:limit1}
\end{eqnarray}
while from the fact that the entrainment frequency cannot be out of the limits $[\min(\dot{\tilde{\theta}}), \max(\dot{\tilde{\theta}})]$ it follows
\begin{eqnarray}
 \min(\tilde{\omega}) - \max|K| < \Omega < \max(\tilde{\omega}) + \max|K|.
\label{eq:limit2}
\end{eqnarray}
Of course, these are broad limits and
only for bounded $\tilde{g}(\tilde{\omega})$ do the boundaries for  $\Omega$  not reach $\pm\infty$.

Hence,  for bounded distributions, as the triangular, and for large enough couplings of opposite signs, the only way the balance equation (\ref{eq:Ensamble_Mean_Final_mix}) can be satisfied, is with the emergence of a TW with a large enough value, so that the  integral $I(K)$ in the same equation will be non-vanishing.
Contrary,  if $I(K)=0$ for the shown example, i.e.
\begin{eqnarray}
\tilde{\omega}-\Omega<|r K|, \ \ \forall \ \tilde{\omega} \ \text{for which} \ \tilde{g}(\tilde{\omega})>0,
\label{eq:condition}
\end{eqnarray}
then the phase balance becomes
\begin{eqnarray}
0=(1-p)/K_1+p/K_2.
\label{eq:condition2}
\end{eqnarray}
If  (\ref{eq:condition2}) holds, then $\Omega$ can have any value that still obeys  condition (\ref{eq:condition}), so each simulation would give a different TW within these limits.
Another consequence from   condition (\ref{eq:condition}) when applied to Eq.~(\ref{eq:Ensamble_Mean_Final}) is that $f_{ens}=0$.

If (\ref{eq:condition2}) is not satisfied, then the oscillators  rearrange, rendering the condition (\ref{eq:condition}) also invalid.
The result is the  appearance of  non-zero $I(K)$ in Eq.~(\ref{eq:Ensamble_Mean_Final_mix}) to impose the balance of the phases.
There may be more than one value for $\Omega$ which impose this balance and if that was the case one might expect multiple solutions or even hysteresis behavior.
In this work however, we only numerically confirm that for bounded natural frequencies, as in the example in Fig.~\ref{fig:4} (a), $\Omega$, when observed in the natural frame, can have two stable values with opposite signs. In other words, numerical realizations could lead to any of the two stable values.
This is illustrated in Fig.~\ref{fig:4} (a) where one such realization is depicted, whilst in different realizations the values on both sides of the mean and the mode of $\tilde{g}(\tilde{\omega})$ appeared for $p$ around  $(0.35, 0.7)$.

For  the same triangular natural frequencies,  if $p>0.65$, then a state similar to the $\pi$ state in the symmetric  model in Section \ref{subsec:TWCon} is observed.
Here, the equality of $\Omega$ and $\tilde{f}_{ens}$ follows from the bounded distribution of the oscillators' natural frequencies, which all become entrained.
Similarly the  integral $I(K)$ vanishes, and the requirement for the phase balance  holds only if $\Omega=0$.
In this way only the first integral of the equation (\ref{eq:Ensamble_Mean_Final_mix})  survives and it gives $<\tilde{\omega}>$, which we set to be $0$ (although  in Fig.~\ref{fig:4} the plots are for non-zero mean frequency distributions, such that the plot (a) has mean $0.8$).

If the natural frequencies are unbounded then $I(K)\neq0$, meaning that a TW  always emerges.
For the simulations we have performed, as seen in Figs.~\ref{fig:4} (b-c), the system was always setting the same value for $\Omega$ that solves Eq.~(\ref{eq:Ensamble_Mean_Final_mix}) for the given parameter range.
Thus, for the log-normally distributed $\omega$, $\Omega$ is always positive, although the existence of another stable TW that also  fulfills the phase balance is not excluded.
The entrainment of all  oscillators is never achieved in this scenario,  and the mean frequencies will never reach  $<\tilde{\omega}>$.
Still they will approach this value when the number of entrained oscillators is increased, either  by increasing the number of conformists, Fig.~\ref{fig:4} (b), or by changing the coupling strength of one of the groups, as shown in  Fig.~\ref{fig:4} (c).
The same plots also show that even when the couplings are all positive, the group frequencies behave qualitatively differently from the case with constant $K$.
This characteristic becomes less obvious once the modes are closer to each other.
To check the generality of the analysis,  we also explored the case when $K$ is bimodal Gaussian distributed.
The results did not show any qualitative difference from the case with bimodal-$\delta$.

\begin{figure}[t!]
\centering
\includegraphics[width=0.22\textwidth]{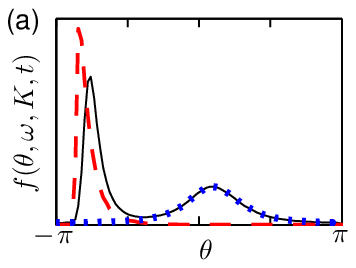} \ \ \ \
\includegraphics[width=0.22\textwidth]{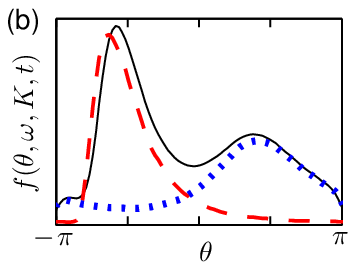} \\
\centering
\includegraphics[width=0.22\textwidth]{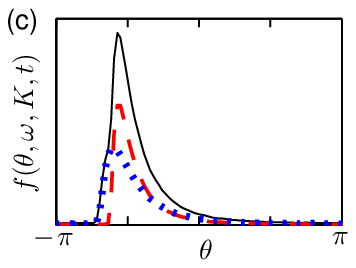} \ \ \ \
\includegraphics[width=0.22\textwidth]{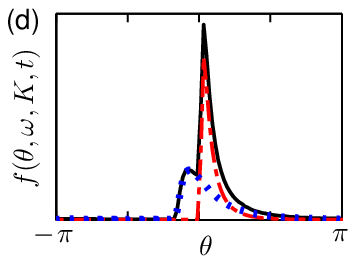}
\caption{(color online) PDF of phases for different states of the ensemble with log-normally distributed natural frequencies and bimodal-$\delta$ distributed coupling strengths.
Oscillators with coupling  $K_1$ are shown with a dashed line (red), those with $K_2$ are following the dotted line (blue,) while the joint distribution is shown with a solid  line (black).
Parameters:
$\mu=0$, $\sigma=1$ and $p=0.5$.
(a) $K_1=-10$ $K_2=20$. Two peaks at a distance smaller than $\pi$ and a TW occurs.
(b) $K_1=-4$ $K_2=8$. Two peaks at a distance $\pi$.
(c) $K_1=2$ $K_2=3$. The peaks merge into a single peak in the joint distribution.
(d) $K_1=2$ $K_2=6$. Two closely separated peaks.
}
\label{fig:5}
\end{figure}

It is expected that for multimodal couplings, clustering emerges and plays a crucial role in defining the values of $\tilde{f}_{ens}$ and $\Omega$. A similar phenomenon was also reported in \cite{Paissan2007} for distributed positive and unimodal $K$, but as a finite-size effect only.
That is to say, the distribution of the locked oscillators  Eq.~(\ref{eqn:Disribution_Locked}) for multimodal-$\delta$ $\Gamma(K)$ will have  peaks on different angles $\psi_i=\arcsin\frac{\omega}{K_i r}$ for each mode $K_i$.
Further evidence for this explanation can be seen in  numerical simulations performed over different $\Gamma(K)$ and $g(\omega)$.
Fig.~\ref{fig:5} presents obtained phase distributions.
As can be seen  three qualitatively different types were identified.

Firstly for very large coupling strengths compared to the width of $g(\omega)$, and for bimodal $\Gamma(K)$ with modes having different signs,  there are two separated peaks corresponding to the contrarians and conformists.
The distance between them is smaller than $\pi$ and the observed TW is similar as in Section \ref{subsec:TWCon}, although the asymmetry in $g(\omega)$ causes asymmetry in the distribution $f(\theta, \omega, K, t)$.
This is shown in Fig.~\ref{fig:5} (a), and also in examples (a-b) of Fig.~\ref{fig:4}, in the range for  $p$ around $0.4-0.65$.

The second scenario differs from the first only in the sense that the  absolute values of the coupling strengths are not much bigger than the width of the natural frequencies. As expected, this corresponds to the $\pi$ state from \cite{Strogatz2011}, but the form of $g(\omega)$ will be mapped onto the peaks which are separated by $\pi$. The asymmetry additionally will influence the values of the $\tilde{f}_{ens}$ and $\Omega$. However, the latter will always be in the interval between the mode and the mean of $\tilde{g}(\tilde{\omega})$.
This means that qualitatively this corresponds to the case with constant $K$ described in Section \ref{subsec:AsF}, but with non-zero values for $\tilde{f}_{ens}$ because of the phase balance equation (\ref{eq:Ensamble_Mean_Final_mix}).

Lastly, if the  couplings distribution have  positive peaks only, depending on their distance apart, the phases will have either one or two close peaks, while still keeping the form of  $g(\omega)$.
Hence,  the case with one peak will be the same as having unimodally distributed $K$ and numerical results suggest that $\tilde{f}_{ens}\approx\langle\tilde{\omega}\rangle$ as for constant $K$, although this is not immediately  obvious from Eqs.~(\ref{eq:Ensamble_Mean_Final}) and (\ref{eq:Ensamble_Mean_Final_mix}).
Finally, if $f(\theta, \omega, K, t)$ is no longer unimodal, then  $\tilde{f}_{ens}$ has values that differ from  $<\tilde{\omega}>$.

The number and the nature of the emerging clusters is not so straightforward for non-$\delta$ $\Gamma(K)$ distributions.
Still, from the PDF of the phases it is clear that for negative modes in $\Gamma(K)$, the peaks in the PDF corresponding to those oscillators do not have the form of $g(\omega)$, but it is more symmetric.
For positive modes, the peaks keep mapping the distribution of the natural frequencies.
This is also an interesting peculiarity  that requires further attention.

The bottom line is that  different and on first sight counterintuitive values for the entrainment and mean ensemble frequency are obtained even for all positive but multimodally  distributed couplings, as long as  the natural frequencies are nonsymmetric.
Nevertheless, a thorough analysis is still needed for this general case of the KM, with the emphasis put on the entrainment at different clusters and their influence  to the  group behavior.

\subsection{Phase shifted coupling function}
\label{subsec:alpha}

\begin{figure}[t!]
\includegraphics[width=0.46\textwidth]{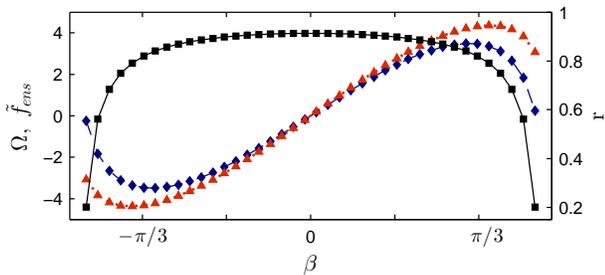}
\caption{(color online) The mean field's amplitude $r$, the phase velocity $\Omega$, and the mean frequency of the ensemble $\tilde{f}_{ens}$ versus the phase shift  $\beta$.
Results from the theoretical predictions for the low-dimensional dynamics, Eqs.~(\ref{eqn:Sakaguchi2} - \ref{eqn:Sakaguchi4}),  are given with a solid  line (black) for $r$, a dotted line (red)  for $\Omega$ and a dashed line (blue) for   $\tilde{f}_{ens}$.
Results from  numerical simulations,  Eq.~(\ref{eqn:KM}), of the ensemble are shown with squares (black) for $r$,  triangles (red) for $\Omega$ and diamonds (blue) for $\tilde{f}_{ens}$.
Parameters:  $\gamma=0.5$,  $K=6$  and $\alpha \in [-4 \pi/9, 4 \pi/9]$.
}
\label{fig:6}
\end{figure}

The final  model  leading to TW of the macroscopic parameters being analyzed here  is actually the earliest generalization  of the KM \cite{Sakaguchi1986}.
This model was introduced to allow for entrainment in frames other than the natural.
The so called Kuramoto-Sakaguchi model has the form
\begin{eqnarray*}
\dot{\tilde{\theta}} = \tilde{\omega} -\, K\, r\sin(\tilde{\theta}_i - \tilde{\psi} - \beta),
 \label{eqn:Sakaguchi}
\end{eqnarray*}
where $\beta$ is a phase shift of the coupling function.
Low-dimensional dynamics of this model with Lorentzian $\tilde{g}(\tilde{\omega})$ can be also obtained using the ansatz \cite{Ott1}, with derivations being similar as for the problem in Section \ref{subsec:TWCon}. It yields
\begin{eqnarray*}
\label{eqn:Sakaguchi2}
\frac{\partial \tilde{z}}{\partial t} + i (\hat{\omega}- i \gamma) \tilde{z} + \frac{K}{2}(\tilde{z} \tilde{\alpha}^2 e^{i \beta}  - \tilde{z}^{\ast} e^{- i \beta}) = 0,
\end{eqnarray*}
which was also obtained in \cite{Pikovsky2011_physD}.
From here, one immediately obtains the long-term evolution of $\psi$, i.e. the frequency of the TW
\begin{eqnarray}
\label{eqn:Sakaguchi3}
\frac{d \tilde{\psi}}{d t} \equiv \Omega = \hat{\omega} + K \sin \beta - \gamma \tan \beta.
\end{eqnarray}
For the mean ensemble frequency in this model,  Eq.~(\ref{eq:Ensamble_Mean_Final}) can be analytically solved and with further substituting stationary values for $r$ and $\Omega$, it transforms to
$$ \tilde{f}_{ens} = \hat{\omega} -sign(\Omega) \gamma \tan \beta.$$
This directly leads to
\begin{eqnarray}
\label{eqn:Sakaguchi4}
 \tilde{f}_{ens} = \hat{\omega} + K \sin \beta - 2 \gamma \tan \beta,
\end{eqnarray}
which can also be straightforwardly obtained from Eq.~(\ref{eqn:MFr4}) after applying the residue theorem for the  integral over $\omega$.
The stationary values of the macroscopic parameters of this system are illustrated  in  Fig.~\ref{fig:6}.
There, it is also obvious that $|\tilde{f}_{ens}|<|\Omega|$ and that the frequencies have odd symmetry, features which are a direct consequence of Eqs.~(\ref{eqn:Sakaguchi3}) and (\ref{eqn:Sakaguchi4}).

It is worth noting that by taking the derivative over $\beta$  from expressions  (\ref{eqn:Sakaguchi3}, \ref{eqn:Sakaguchi4}), extreme values for both mean frequencies follow as
$$\beta_{max/min} = \pm \arccos \sqrt[3]{2 \gamma / K},$$ for $\Omega$ and
$$\beta_{max/min} = \pm \arccos \sqrt[3]{\gamma / K},$$ for $\tilde{f}_{ens}$.
Of course, since the population should  be  coherent, parameters have to be chosen such that  $K>K_c=2 \gamma / \cos \beta$ holds.
Thus, the values of the phase shift $\beta$ that produce maximum deviation in  both mean ensemble and mean field frequencies are directly obtained.

\section{Summary}
\label{sec:disc}

The TW state in the KM, occurs whenever the symmetry in either the natural frequencies or the coupling function itself is broken.
For asymmetric coupling strengths on the other hand, the wave occurs only for certain coupling parameters.
The results of this study indicate that in populations with an asymmetric, bell-shaped distribution of the non-synchronized oscillators and equal couplings, the mean ensemble frequency is always the mean of the natural frequencies.
In contrast, the TW of the synchronized oscillators -- the mean field frequency -- has a value between the mean and the mode of those.
When the asymmetry originates from the coupling strengths or the coupling function, both mean frequencies have non-trivial values.
In particular they  differ from the mean or the modes of the frequency distribution.

The case of the asymmetric unimodal frequency distribution has quite a straightforward explanation.
Here the frequency of synchronization is a result of the interplay between the cluster of the  locked oscillators and the whole ensemble.
Synchronized oscillators tend for frequency locking at the mode, where the density of similar oscillators is the highest, whilst the
influence of all oscillators would be balanced if it is at the mean of all natural frequencies.
The compromise is achieved through the self-consistent system  for the phase balance and Eq.~(\ref{eq:Ensamble_Mean_Final_AS}) follows directly from this interplay.
This balance would be destroyed if the nucleus is either on the peak or on the mean.

By increasing  the number of locked oscillators, the mean field frequency shifts towards the less skewed side of the distribution, i.e. towards the mean, because the number and the influence of the drifting oscillators decreases.
Finally, once all the oscillators get locked, the frequency of the entrainment will be exactly the mean of the distribution.
In this way, the influence of all oscillators is equal.
The same explanation can be applied for bimodal or multimodal cases, since again there is only one cluster of synchronized oscillators in the TW state.

On the other hand, in the case of contrarians and conformists, the physical explanation of the mean ensemble frequency, the frequency of entrainment and their difference, is not yet  very clear. Previous work went  as far as showing that in some parameter spaces for opposite sign couplings, both peaks in the distribution of the phases lose their stability. Instead, they  start to chase each other and are no longer a distance of $\pi$ apart, hence producing the TW.
Here we have emphasized that this behavior simply acts to preserve  the phase balance, which directly gives the value of the entrainment frequency, while its influence on arranging the non-synchronized oscillators also makes it responsible for the mean ensemble frequency.
Despite this, we believe that even by showing, analytically and numerically, that the velocity of the mean phase and the frequency of the ensemble  differ in this case, an important characteristic that seems to be unique for TW states only  is immediately  revealed.
This explanation was missing when the model was first introduced \cite{Strogatz2011}, and clear distinction between macroscopic frequencies was not made.
Moreover, ``the mean phase velocity'' obtained through the low-dimensional dynamics was defined as the mean ensemble frequency.

In the more general and  realistic situation, with asymmetric frequencies and distributed couplings,   the  macroscopic frequencies can have values in very unexpected boundaries -- even outside the limits of the natural frequencies.
The reason for this complex behavior is again a result of the system's self-arranging, such that the influence of the non-synchronized oscillators to the mean field amplitude vanishes and the phase balance is maintained.
In order for this to be achieved for multimodally distributed couplings, there are different clusters of synchronized oscillators for each mode.
This case also shows possibilities for further research either in eventual hysteresis dynamics and seeking the number of stable solutions, or in the clustering phenomena for distributed coupling strengths.

When the  asymmetry is induced through the coupling function itself the macroscopic frequencies always differ, in a similar way to the case when it stems from the coupling parameters.
In that sense the mean ensemble frequency is of the same sign, but  with lower absolute value than the mean field frequency.
Also,  their dependance on the phase shift  is nonlinear, and the ways they respond to the shift do not coincide, with both of them having the highest values for different phase shifts.

Taken together, these results suggest that whenever the population is experiencing a TW, the locking of the oscillators is at a frequency different to the mean of the instantaneous frequencies.
Since in inverse problems or in experiments it is often only the macroscopic parameters that can be obtained, a clear  interpretation  of the observed \textit{mean frequency} is always needed.
Moreover, the asymmetric scenarios tend to be far more abundant in the real physical systems.
That is to say, asymmetry of the natural frequencies can immediately make any scenario more realistic, whilst few of the examples with opposite coupling strengths can be traced in inhibitory and excitatory neurons \cite{WilsonCowan} or in social dynamics \cite{Lama}.
As for the phase-shifted coupling function, although it was introduced to describe the formation of nonlinear waves in non-oscillatory media \cite{Sakaguchi1986}, it can be used to model various phenomena, such  as Josephson junctions \cite{Wiesenfeld1996}, mammalian intestines and heart cells \cite{Winfree1980}.
Hence, in the models based on measurements, like those describing the brain dynamics \cite{Jane1}, one should precisely define to which of the macroscopic frequencies, of the mean field or  mean ensemble, the measured frequency corresponds.
Finally, this work reveals the strong need for future research that should explain the physical link between the observed mean frequency of any system with cooperative dynamics and the two macroscopic frequency parameters described here for non-trivial cases.

\section*{Acknowledgments}

The study was supported by the EPSRC, SP is supported by the Lancaster University PhD grant and LB was  supported by Faculty of Computer Science and Engineering at the SS.Cyril and Methodius University and  ERC Grant $\#$ 266722 (SUMO project).
We are grateful to Gemma Lancaster and Phil Clemson for their useful comments on the manuscript.

\appendix

\section{General formula for the mean ensemble frequency}
\label{sec:app1}

We work in the reference frame rotating with $\Omega=\dot{\tilde{\psi}}$.
In this frame, the locked oscillators are frozen and one has to consider the drifting ones only, Eq.~(\ref{eqn:Disribution_Unlocked}).
This can be applied in the expression for the mean frequency of the ensemble, Eq.~(\ref{eqn:MFr2}), which becomes
\begin{widetext}
\begin{eqnarray}
f_{ens} &=& \int_{-\pi}^{\pi} \int\int_{|\omega|>|Kr|}(\omega - K r \sin\theta) g(\omega)\Gamma(K)\frac{\sqrt{(\omega)^2-(Kr)^2}}{2\pi|\omega-Kr\sin(\theta)|} d\omega dK d\theta = I_1-I_2.
\label{eq:Ensemlbe Mean_I1_and_2}
\end{eqnarray}
\end{widetext}
The first integral is
\begin{eqnarray}
I_1 = \int \int_{|\omega|>|Kr|} \omega \Gamma(K) g(\omega) dK d\omega,
\label{eq:Ensemlbe Mean_I1}
\end{eqnarray}
because of the probability normalization
\begin{equation}
\int_{0}^{2\pi} \frac{\sqrt{(\omega)^2-(Kr)^2}}{2\pi|\omega-Kr\sin\theta|}d\theta = 1.
\end{equation}
The second integral requires more calculation.
Integrating over the phases first, for positive frequencies $\omega>|Kr|$ it yields
\begin{equation}
\int_{0}^{2\pi} \frac{\sqrt{\omega^2-(Kr)^2}Kr\sin\theta}{2\pi|\omega-Kr\sin\theta|} d\theta =   \omega - \sqrt{\omega^2-(Kr)^2},
\label{eqn:I1}
\end{equation}
while for negative frequencies $\omega<-|Kr|$
\begin{equation}
\int_{0}^{2\pi} \frac{\sqrt{\omega^2-(Kr)^2}Kr\sin\theta}{2\pi|\omega-Kr\sin\theta|} d\theta =  \omega + \sqrt{\omega^2-(Kr)^2}.
\label{eqn:I2}
\end{equation}
It is interesting to note that both integrals are even in $K$.
Following this, the second integral, $I_2$, simply becomes
\begin{eqnarray}
&&I_2 =  \int_{-\infty}^{\infty} \int_{-\infty}^{-|Kr|}  g(\omega) \Gamma(K)  [\omega + \sqrt{\omega^2-(Kr)^2}] d\omega dK  \nonumber\\
 + && \int_{-\infty}^{\infty} \int_{|Kr|}^{\infty} g(\omega) \Gamma(K)  [\omega - \sqrt{\omega^2-(Kr)^2}] d\omega dK.
\label{eq:Ensemlbe Mean_I2_Solved}
\end{eqnarray}
After partial cancelation of $I_2$  with $I_1$  because of (\ref{eq:Ensemlbe Mean_I1_and_2}) one obtains a simple formula for calculation of the mean ensemble frequency in the reference frame of the order parameter
\begin{widetext}
\begin{eqnarray}
f_{ens} &=& \int_{|Kr|}^{\infty} \int_{-\infty}^{\infty} g(\omega) \Gamma(K) \sqrt{\omega^2-(Kr)^2}d\omega dK
- \int_{-\infty}^{-|Kr|} \int_{-\infty}^{\infty} g(\omega) \Gamma(K) \sqrt{\omega^2-(Kr)^2} d\omega dK \nonumber\\
&=&\int_{|Kr|}^{\infty} \int_{-\infty}^{\infty} [g(\omega)-g(-\omega)] \Gamma(K)\sqrt{\omega^2-(Kr)^2}d\omega dK.
\label{eq:Ensamble_Mean_FinalA}
\end{eqnarray}
Note that $g(\omega)$ is centered in $-\Omega$ and hence is asymmetric, even for symmetric $\tilde{g}(\tilde{\omega})$ which is often the case.
Returning to the original frequencies $\tilde{\omega}$, the last expression becomes
\begin{eqnarray}
f_{ens} = \int_{|Kr|}^{\infty} \int_{-\infty}^{\infty} [\tilde{g}(\omega+\Omega)-\tilde{g}(-\omega+\Omega)] \Gamma(K)\sqrt{\omega^2-(Kr)^2}d\omega dK.
\label{eq:Ensamble_Mean_FinalB}
\end{eqnarray}
\end{widetext}

\bibliography{apssamp}

\end{document}